\documentclass[12pt]{article}

\usepackage[utf8]{inputenc}
\usepackage[T1]{fontenc}
\usepackage{amsmath, amssymb, amsthm}
\usepackage{graphicx}
\usepackage{booktabs}
\usepackage{hyperref}
\usepackage[round]{natbib}
\usepackage{geometry}
\usepackage{setspace}
\usepackage{caption}
\usepackage{float}

\title{Two-Part Forecasting for Time-Shifted Metrics}
\author{
  Harrison Katz\thanks{Corresponding author. Email: \texttt{harrison.katz@airbnb.com}} \\
  \textit{Airbnb, Inc.}
  \and
  Erica Savage \\
  \textit{Airbnb, Inc.}
  \and
  Kai Thomas Brusch \\
  \textit{Airbnb, Inc.}
}
\date{}

\begin{document}

\maketitle

\begin{abstract}
Many commercial sectors (such as hospitality) face the challenge of forecasting metrics that span multiple time axes, where the timing of an event's occurrence is distinct from the timing of its recording or initiation. We present a novel two-part forecasting methodology that addresses this challenge by treating the forecasting process as a time-shift operator. The approach combines univariate time series forecasting to predict total bookings on booking dates with the Bayesian Dirichlet Auto-Regressive Moving Average (B-DARMA) model. The aim is to forecast the allocation of future bookings across different trip dates based on the time between booking and trip (lead time). This approach provides a sensible solution for forecasting demand across different time axes, offering interpretable results, flexibility, and the potential for improved accuracy. The efficacy of the two-part methodology is illustrated through an analysis of Airbnb booking data.
\end{abstract}

\noindent\textbf{Keywords:} time series forecasting, compositional data, Bayesian methods, Dirichlet distribution, lead time, hospitality

\section{Introduction}\label{sec:intro}

Accurate demand forecasting is essential across industries for optimizing operations, managing resources, and strategic planning. Yet many sectors face the challenge of \textit{time-shifted metrics}, where the event date (e.g., booking, order, or trade date) does not match the date the service is consumed or settled (trip date, delivery date, settlement date). Such temporal separation often causes inconsistencies and reduced accuracy when traditional, single-axis forecasting methods are applied.

\begin{figure}[H]
  \centering
   \includegraphics[width=0.7\textwidth]{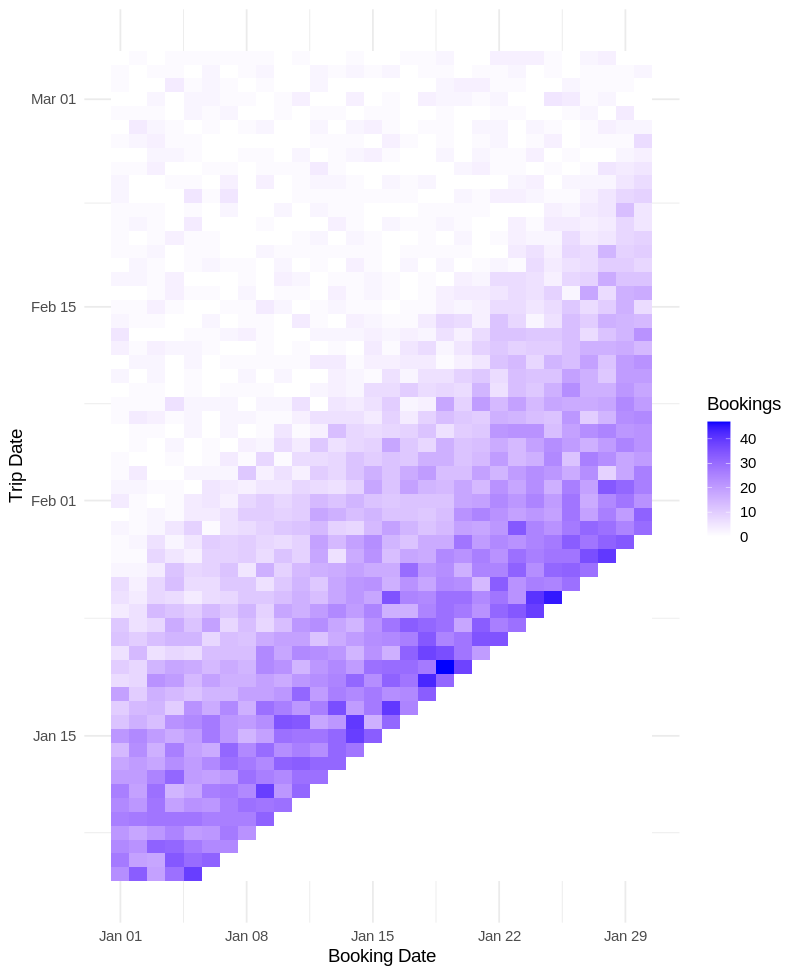}
  \caption{Heatmap of simulated booking counts for each (booking date, trip date) pair.}
  \label{fig:heatmap}
\end{figure}

Figure~\ref{fig:heatmap} illustrates the time-shifted nature of the data, where a single booking date might correspond to a trip starting immediately or up to weeks later. In hospitality settings, the total nights (or stays) initially forecast from the booking perspective can change by the time the trip date arrives, often due to cancellations or modifications. This time-shifted nature can also appear in supply chain (purchase vs.\ delivery), retail (sale vs.\ shipment), healthcare (appointment vs.\ consultation), and finance (trade vs.\ settlement). Traditional forecasting approaches that operate within a single temporal framework struggle to track how the metric transitions from one origin to another.

While hierarchical time series methods aim to reconcile forecasts at different organizational or product levels \citep{Hyndman2011}, they do not directly handle the ``two-axis'' structure arising from time-shifted metrics. Here, the primary challenge is that one axis (e.g., booking date) must be aligned with a distinct time axis (e.g., trip date), rather than just different levels or categories on a single timeline. Similarly, temporal aggregation approaches address scaling forecasts up or down in time granularity \citep{Silvestrini2008}, but they do not typically involve distributing one metric across another time dimension. Hence, while our framework is conceptually adjacent to these literatures, time-shifted forecasting remains a unique problem requiring distinct methods.

In contrast, compositional data analysis provides a way to model proportions that sum to a whole \citep{Aitchison1986, Zheng2017}, but its application to lead-time distributions in a multi-axis setting has been limited. Hybrid strategies that combine univariate and compositional techniques \citep{Armstrong2001} can help bridge these gaps.

Our paper introduces a two-part forecasting methodology that treats the process as a time-shift operator. We first project total demand on the booking axis, then translate those forecasts to the trip axis using a compositional time series model. For Airbnb, these predictions inform a wide range of decisions, such as dynamic pricing, host recruitment, staffing programs, and marketing campaigns, so that supply can be aligned with expected guest stays. Even a seemingly small decrease in forecast error can translate into substantial cost savings or revenue optimizations at scale. After describing this approach in more detail, we present our analysis of Airbnb data and conclude with broader insights on how it can apply to other industries.

\section{Methodology}\label{sec:methodology}

Full mathematical derivations of the Bayesian Dirichlet Auto-Regressive Moving Average (B-DARMA) model, including the additive log-ratio transformations, can be found in our online supplement. \citet{Katz2024,forecast7030032,forecast7040062} provides additional technical details on the methodology and our analysis.

The two-part model structure begins with a forecast of total bookings made on a booking date regardless of trip date. For this, a univariate time series model (e.g., ARIMA, Prophet, or exponential smoothing) is applied to historical daily bookings. Given its robustness to trend changes and seasonality, we used Prophet \citep{Taylor2018} to obtain our forecast of the total bookings for each future date.

The second part of the model uses B-DARMA for lead-time allocations. It begins by creating a vector representing the proportions of bookings made on each day that fall into each lead-time bucket (e.g., 0--1 month, 1--2 months, etc.). Then we utilize B-DARMA, which provides a structure for the means in the transformed space.

Having computed the forecasts of total bookings on each date, along with the proportions of bookings each day falling into each lead-time bucket, we multiply them. This combines the two parts of the model structure.

\section{Data}\label{sec:data}

Our example employs two anonymized Airbnb datasets:

\begin{itemize}
  \item \textbf{City A:} A large metropolitan market with strong seasonal variability.
  \item \textbf{City B:} A midsized leisure destination with more moderate seasonality.
\end{itemize}

\begin{figure}[H]
  \centering
  \includegraphics[width=0.9\textwidth]{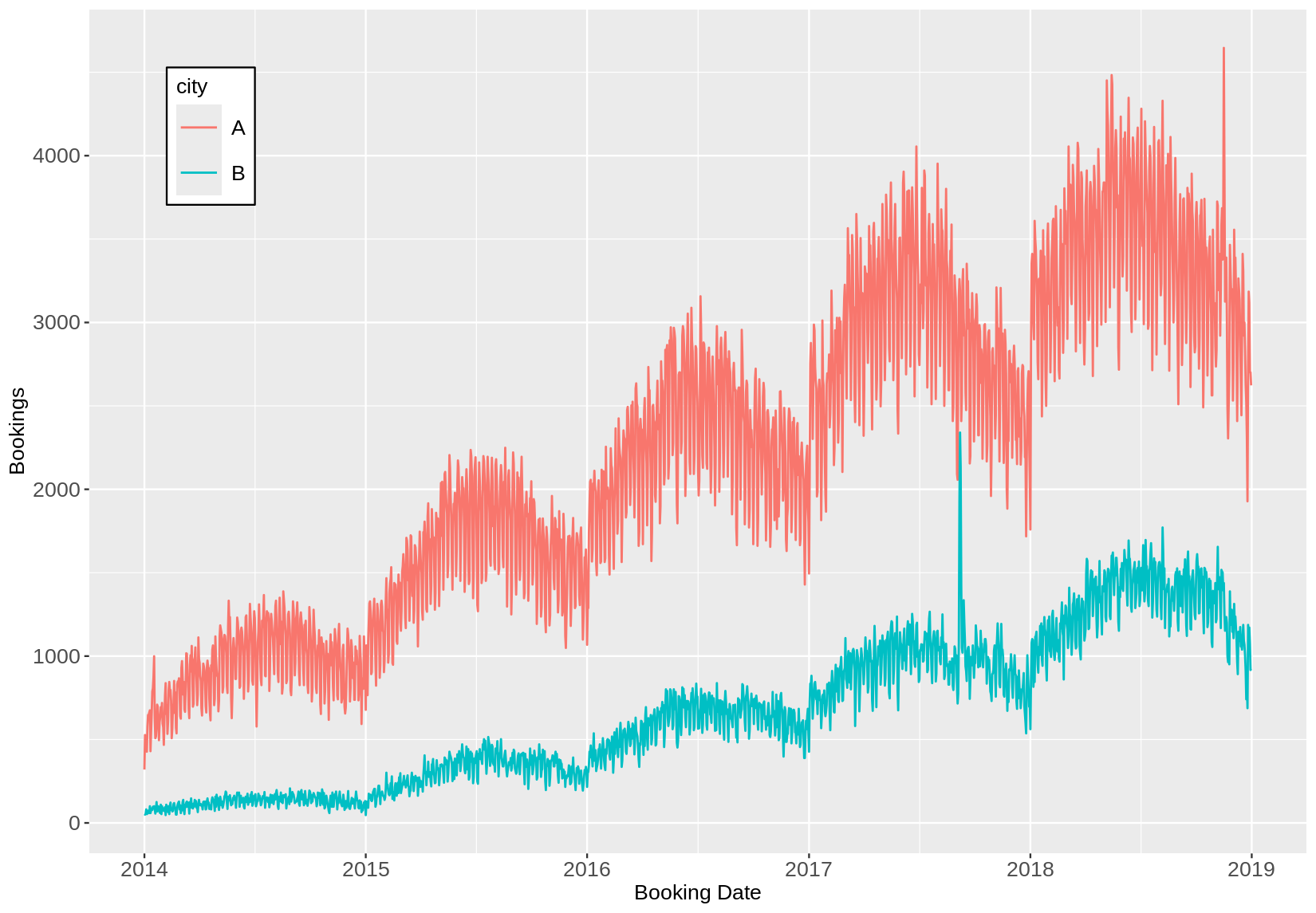}
  \caption{Daily Airbnb bookings by booking date over the training period (2014--2018).}
  \label{fig:daily_bookings}
\end{figure}

\begin{figure}[H]
  \centering
   \includegraphics[width=0.9\textwidth]{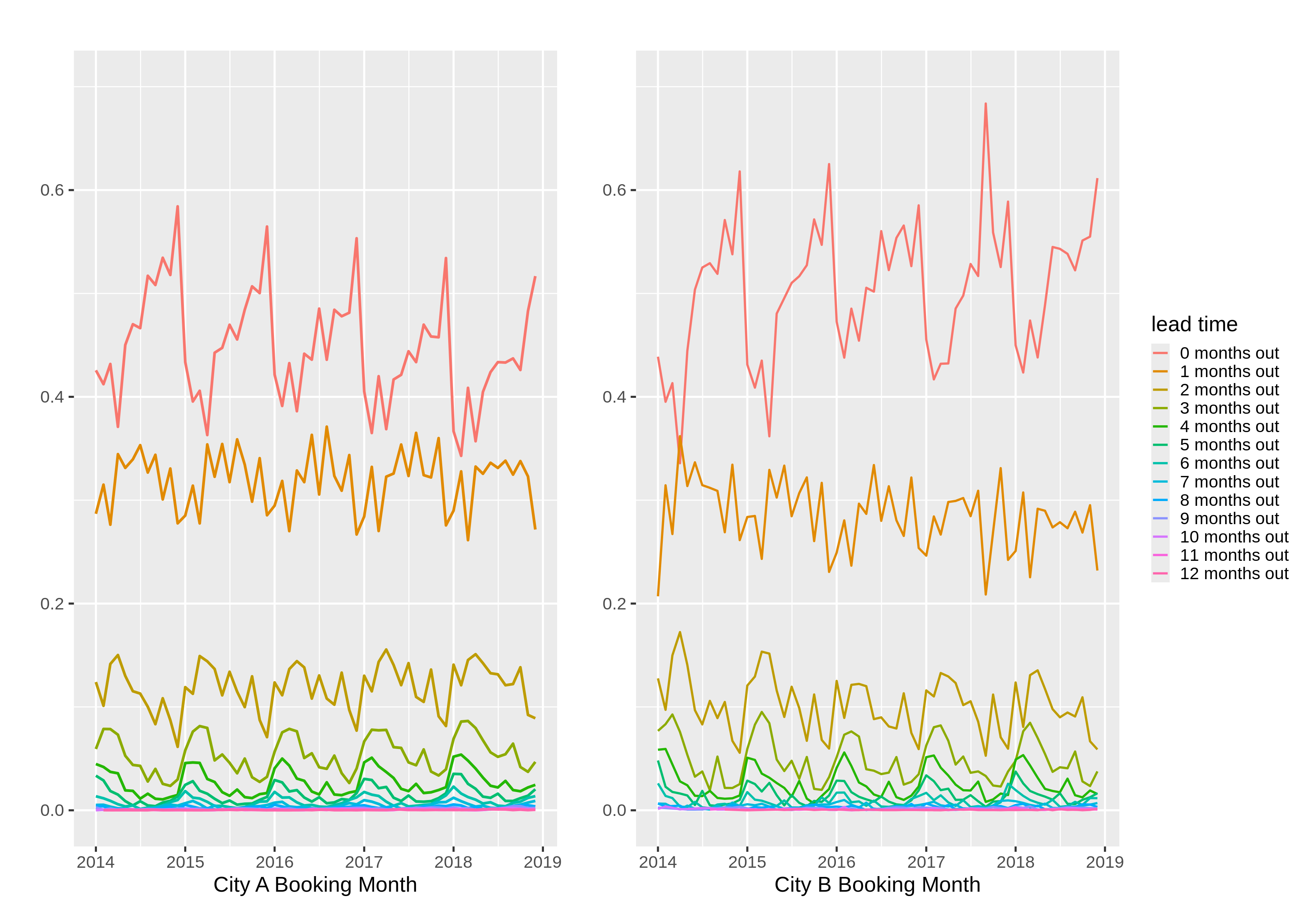}
  \caption{Proportion of monthly bookings by lead time over the training period (2014--2018).}
  \label{fig:leadtime_props}
\end{figure}

Each dataset spans six full years (January 2014 to December 2019, before the COVID lockdowns) at a daily granularity. Each contains the number of bookings made on day $t$, the trip date (or month) of each booking, and lead time in months. We create monthly lead-time buckets from 0 to 12, forming 13 possible lead times. The daily bookings are shown in Figure~\ref{fig:daily_bookings} while the lead-time proportions are shown in Figure~\ref{fig:leadtime_props}.

In City B, we observe a notable spike on the booking-date axis around September 1, 2017. In City A, there is a lesser spike around November 2018. These surges are not due to data errors; instead, they likely stem from external triggers that prompted many reservations within a short window. Typical examples include extreme weather advisories leading to last-minute changes, sporting events where the final location is confirmed late in a playoff series, or major concerts/music festivals that announce dates and release tickets at once, causing a rapid influx of bookings when fans learn the time and venue. Such real-world events can produce abrupt jumps in the booking-date series, even if the trips themselves occur on future dates.

Our training period was five full years (2014 through 2018) with test period January 1 through December 31 of 2019. We fit all models on the training window, then compare forecast accuracy over the test window (one full calendar year).

\section{Two-Part Method Implementation}\label{sec:implementation}

In our approach, we first forecast total daily bookings on the booking-date axis for each city by applying Prophet \citep{Taylor2018} to the five years of training data (2014--2018). For City A, the model incorporates weekly and annual seasonal components, as well as holiday factors reflective of a large metropolitan market. City B, by contrast, is a midsized leisure destination and therefore employs slightly different holiday and event indicators to account for its unique patterns. From these daily forecasts, we produce monthly total bookings for the one-year test window.

Having obtained the expected number of bookings for each day, we then address how those bookings spread across various lead times. Specifically, we record the proportion of bookings, ranging from zero to 12 months in advance, on each booking month. To model these allocations, we use a B-DARMA(1,0) approach in which this month's lead-time distribution depends on recent patterns. We also incorporate monthly seasonality via Fourier terms and include a linear trend to capture shifting booking behaviors throughout the calendar. This yields a monthly sequence of lead-time proportions for 2019, providing insight into how the share of last-minute versus long-term bookings evolves over time.

With total daily bookings and monthly lead-time proportions in hand, we first aggregate our daily booking-date forecasts into monthly totals. We then multiply each month's total bookings by the corresponding lead-time proportions and shift these results forward by the appropriate monthly offset (0 to 12). By summing across all booking months that align with a given trip month, we obtain the final forecast of how many stays (or similar events) will occur in that month. This step effectively translates the original booking-date perspective into the trip-date perspective, revealing when actual consumption or usage is expected to take place.

\section{Benchmark: Bottom-Up Prophet Approach}\label{sec:benchmark}

To highlight the potential benefits of modeling lead times compositionally, we compare our method with a simpler bottom-up Prophet approach. In this benchmark, we create a separate univariate Prophet forecast for each monthly lead-time bucket. For instance, one model predicts the number of bookings on day $t$ that are for a trip starting within the same month, another model does so for bookings with a trip date that falls in the following month, and so on up to 12 months out from the booking month. Summing these daily forecasts across all buckets gives an estimate of total daily bookings, which we can compare to the original Prophet forecast in our Part~1. Converting each bucket's forecast to a proportion of the monthly total also provides a compositional perspective we can pit against the B-DARMA outputs.

In the supplementary materials, we also provide results from a benchmark that applies Prophet at the monthly level (by lead-time bucket), allowing us to compare performance under a coarser temporal aggregation.

\section{Metrics and Results}\label{sec:results}

We assess both the booking-date and trip-date forecasts using several criteria. MAPE captures how far off the forecast is in relative percentage terms, while MAE shows absolute differences between forecasted and actual totals. We also include a normalized $L^1$ distance for compositional vectors, sometimes called Manhattan distance, to measure how closely the forecasted proportions align with actual lead-time distributions for each booking month \citet{KATZ2025100185}. In situations where the specific breakdown of bookings across lead times is critical, such as staffing hotels or planning supply chains, this compositional accuracy can be as important as the raw total demand forecast.

\subsection{Booking-Date Axis}

Table~\ref{tab:results} shows monthly aggregated forecast errors on the booking-date axis for 2019. It compares our two-part method to the bottom-up Prophet approach over the test window, in terms of both MAE and MAPE. It also shows mean normalized $L^1$ distance for lead-time distributions to the trip-date axis. Lower values in each metric indicate better performance.

\begin{table}[H]
  \centering
  \caption{Performance metrics over the test window (2019).}
  \label{tab:results}
  \begin{tabular}{llrrr}
    \toprule
    City & Method & Booking Date MAE & Booking Date MAPE & Lead-Time Mean Norm.\ $L^1$ \\
    \midrule
    A & Two-Part    & 5{,}083 & 4.80\% & 0.0229 \\
    A & Bottom-Up   & 5{,}336 & 5.07\% & 0.0389 \\
    B & Two-Part    & 1{,}406 & 3.07\% & 0.0300 \\
    B & Bottom-Up   & 1{,}455 & 3.15\% & 0.0499 \\
    \bottomrule
  \end{tabular}
\end{table}

Additionally, we tested a Prophet-based benchmark at the monthly level (rather than daily) by lead-time bucket; the supplementary materials provide full details. Results were largely consistent with the daily bottom-up forecasts once aggregated to a monthly scale.

For both City A and City B, the two-part approach (Part 1 total + lead-time from B-DARMA) generally tracks overall daily bookings well, with average MAPE around 4.8\% for City A and 3.1\% for City B. The bottom-up Prophet sum occasionally lags behind changes in overall level demand, having marginally higher MAPE (5.1\% for City A and 3.2\% for City B).

\subsection{Lead-Time Distribution}

\begin{figure}[H]
  \centering
  \includegraphics[width=\textwidth]{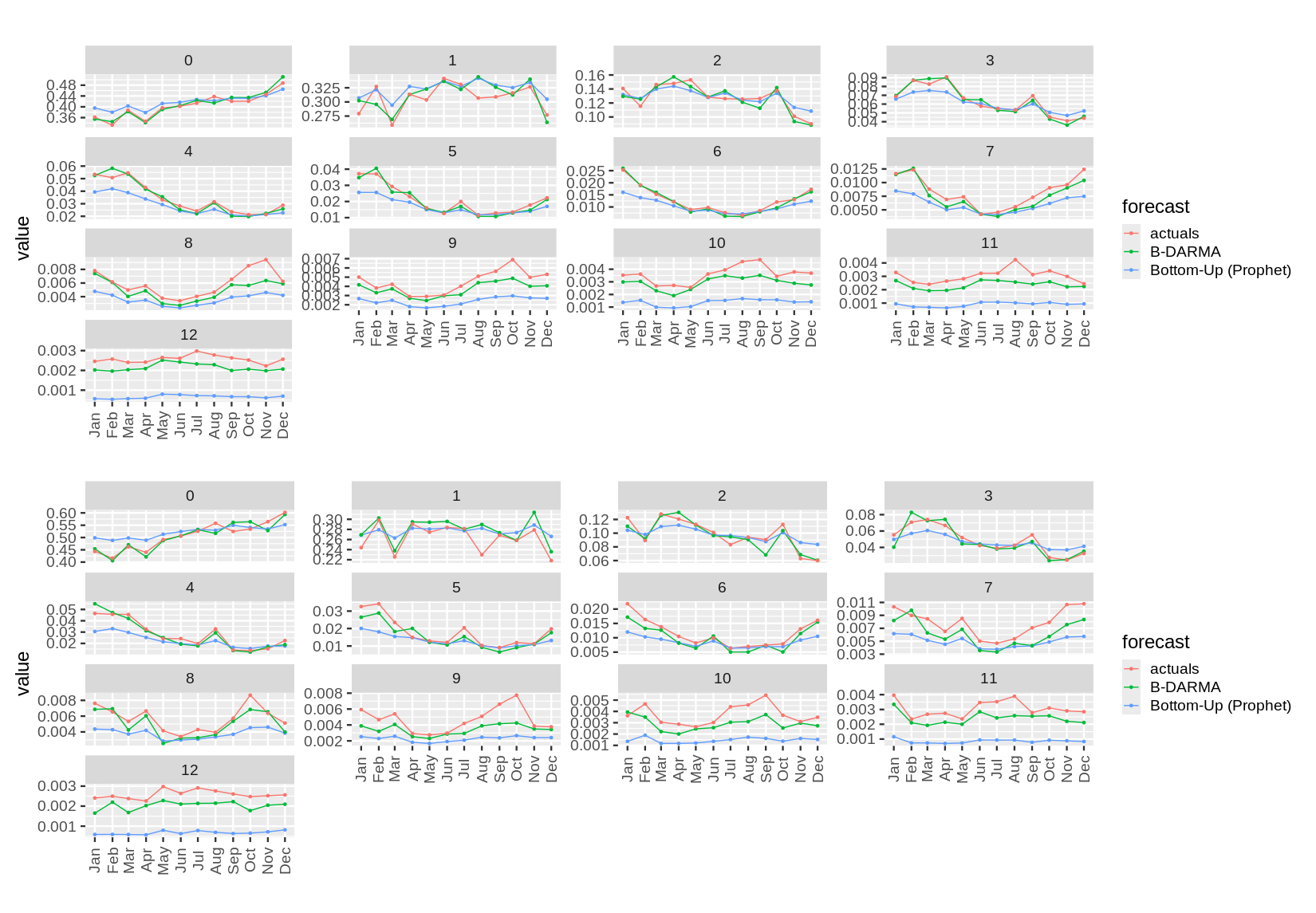}
  \caption{Monthly lead-time proportions in 2019 for Cities A (top) and B (bottom). Each subplot corresponds to one of 13 lead-time buckets. Actual proportions appear in red, with forecasts from B-DARMA in green and bottom-up Prophet in blue.}
  \label{fig:leadtime_forecasts}
\end{figure}

The far-right column in Table~\ref{tab:results} compares the mean normalized $L^1$ distance for the lead-time distributions in Cities A and B. Figure~\ref{fig:leadtime_forecasts} shows the forecasts for the lead times and the normalized $L^1$ for each booking month in the 2019 test window.

\begin{figure}[H]
  \centering
  \includegraphics[width=0.9\textwidth]{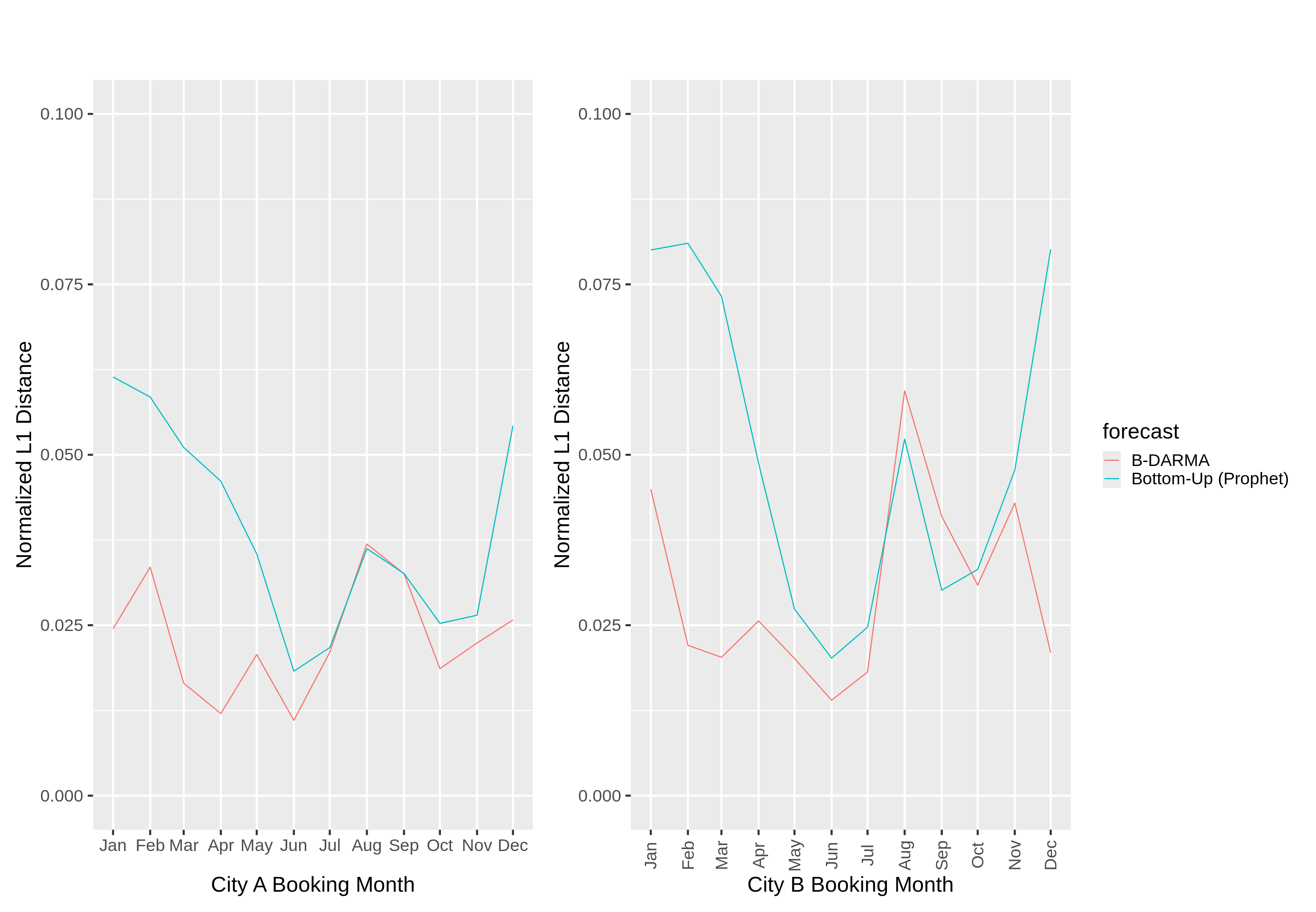}
  \caption{Normalized $L^1$ distance by booking month in 2019 for each forecast method (B-DARMA vs.\ bottom-up Prophet). The left panel shows results for City A, and the right panel for City B. Lower values indicate closer agreement between forecasted and actual lead-time distributions.}
  \label{fig:l1_distance}
\end{figure}

In both markets, the two-part (B-DARMA) approach outperforms the bottom-up Prophet model. For City A, B-DARMA achieves a mean normalized $L^1$ distance of 0.0229 (vs.\ 0.0389). City B, while exhibiting higher overall volatility, shows a similar pattern: 0.030 (B-DARMA) vs.\ 0.0499 (bottom-up) for the $L^1$ distance. These results indicate that the two-part method's compositional framework captures cross-bucket correlations more effectively, leading to more accurate lead-time allocations than independent univariate forecasts.

\section{Discussion}\label{sec:discussion}

Accurate and consistent forecasting across multiple time axes remains a challenge in many industries. By treating the forecasting process as a time-shift operator, our two-part methodology not only provides coherent forecasts but also improves interpretability and flexibility. In our analysis of City A and City B, separating the forecasts into total bookings (Part 1) and compositional lead-time distributions (Part 2) led to lower error rates on both the booking-date and trip-date axes compared to a bottom-up Prophet approach. This is because B-DARMA captures the cross-bucket correlations that univariate bucket-by-bucket models often overlook.

A major advantage of this two-part method lies in its modularity and adaptability. Adjusting total forecasts in response to macroeconomic or event-driven shocks, for instance, can be done without refitting the lead-time model, allowing rapid scenario analyses when unexpected changes occur. This modular structure also extends naturally to short-horizon forecasting, where some future bookings are already known. If, for example, today is January 21 and occupancy is forecast for January 30--31, a portion of those stays may already be booked. In such cases, these existing reservations serve as a baseline or ``backfill'' on the trip date, while the univariate booking-date forecast and B-DARMA compositional vectors project any additional bookings that might still materialize. Because lead-time allocations remain anchored to the booking date, it is straightforward to align incremental forecasts with the appropriate trip dates, preserving both the scenario-testing capability for total demand and the advantage of having prior knowledge about near-term reservations.

Moreover, the B-DARMA model can incorporate exogenous covariates not just for the booking date (e.g., day-of-week, macro factors) but also for the trip date. One could, for instance, add a Super Bowl or Easter indicator to the relevant trip-date bucket, thereby shifting proportions if that event drives higher demand at certain lead times. By adding holiday/event covariates in the compositional (alr) space, the model can directly link a ``trip date'' feature to the observed lead-time allocations, ensuring that special events feed back into both total demand and how that demand is distributed over the booking horizon. This flexibility helps unify the perspective of forecasting ``for'' a particular day (trip date) with the perspective of forecasting ``on'' a particular day (booking date) under one cohesive framework.

However, this methodology is not without limitations. Splitting the forecasting process into two parts may overlook interactions between total demand and lead-time behavior that a unified model could potentially capture. Additionally, B-DARMA's compositional foundation generally assumes that proportions remain strictly positive, so extremely sparse or zero-valued lead-time buckets can pose modeling challenges. If lead times are of no particular interest or remain largely static, the added complexity of a compositional model may not justify its use, and a simpler bottom-up or univariate strategy could suffice. Another potential enhancement is to incorporate probabilistic intervals, given that B-DARMA's Bayesian framework naturally supports credible intervals for lead-time proportions.

Despite these caveats, our results suggest that for scenarios where dynamic lead times meaningfully influence resource allocation or demand planning, a two-part compositional framework delivers valuable improvements in accuracy and interpretability. Extending the approach to hierarchical structures (e.g., city vs.\ region) or further temporal aggregations offers promising avenues for future research.

\section*{Acknowledgements}

The authors thank Sean Wilson, Liz Medina, Jenny Cheng, Jess Needleman, and Peter Coles for helpful discussions, Ellie Mertz for championing the research, and Lauren Mackevich and Lori Callo for their indispensable operational support.

\section*{Data and Code Availability}

The primary dataset used in this study is not publicly available due to confidentiality constraints. However, the supplementary material containing the full model specifications, including the Stan code for the B-DARMA model, is available for public access at our GitHub repository: \url{https://github.com/harrisonekatz/consistent_forecasting_bdarma_paper}.

\bibliographystyle{apalike}
\bibliography{references}

\end{document}